\documentclass{Interspeech}
\interspeechcameraready 
\usepackage{etoolbox}
\usepackage{subfigure}
\usepackage{subcaption}
\usepackage{CJKutf8}
\usepackage{multirow} 
\usepackage{ragged2e} 
\usepackage{booktabs, makecell, multirow, tabularx}
\usepackage{multicol}
\usepackage{footmisc}
\usepackage{setspace}
\usepackage{comment}

\title{Universal Preference-Score-based Pairwise Speech Quality Assessment}

\author{Yu-Fei}{Shi}
\author{Yang}{Ai$^*$}
\author{Zhen-Hua}{Ling}

\affiliation[nocounter]{National Engineering Research Center of Speech and Language Information Processing}{\\University of Science and Technology of China}{Hefei, P. R. China}
\email{zkddsr2023@mail.ustc.edu.cn, \{yangai, zhling\}@ustc.edu.cn}

\begin{document}

\maketitle
\renewcommand{\thefootnote}{\fnsymbol{footnote}}
\footnotetext[1]{Corresponding author. This work was funded by the National Nature Science Foundation of China under Grant U23B2053 and 62301521, and the Anhui Provincial Natural Science Foundation under Grant 2308085QF200.}
\renewcommand{\thefootnote}{\arabic{footnote}}
\begin{abstract}

To compare the performance of two speech generation systems, one of the most effective approaches is estimating the preference score between their generated speech. 
This paper proposes a novel universal preference-score-based pairwise speech quality assessment (UPPSQA) model, aimed at predicting the preference score between paired speech samples to determine which one has better quality. 
The model first predicts the absolute mean opinion score (MOS) for the two speech samples separately, and then aggregates them into a relative preference score using a preference function. 
To address the scarcity of preference data, we also construct a new pairwise speech dataset based on a MOS dataset for experiments. 
Experimental results confirm that, whether in training scenarios with different data types and label conditions, or in both in-domain and out-of-domain test scenarios, the prediction accuracy of UPPSQA outperforms that of the baseline models, demonstrating its universality.
    
\end{abstract}
\vspace{-1.5mm}
\section{Introduction}
\label{Introduction}

Human subjective ratings are the gold standard for assessing speech generation systems such as text-to-speech (TTS) and voice conversion (VC). 
Participants are asked to listen to the speech samples to be evaluated and rate one or more samples based on specific criteria. 
One of the most common subjective evaluation method is the mean opinion score (MOS) testing. 
As subjective MOS testing requires a significant amount of time and funding, several automatic MOS-based speech quality assessment (SQA) models have emerged in recent years, e.g., MOSNet \cite{lo2019mosnet}, LDNet \cite{huang2022ldnet}, SSL-MOS \cite{cooper2022generalization}, UTMOS \cite{saeki2022utmos}, and SAMOS \cite{shi2024samos}. 
These models can provide a MOS for a single speech sample, with SSL-MOS \cite{cooper2022generalization} being the first to leverage the high-level semantic information of SSL models into MOS prediction. 
SAMOS \cite{shi2024samos} builds on SSL-MOS by incorporating richer acoustic information and further advancing system-level MOS prediction accuracy. 

Unlike MOS, which evaluates individual speech samples, preference listening test \cite{shah2014better} focuses on evaluating paired speech samples to determine which one has better quality. 
Preference listening test is widely used in many speech generation tasks because it makes it easier to assess the statistical differences between two systems \cite{vazquez2002reliability,kiritchenko2017best}. 
Therefore, in recent years, automatic preference-score-based SQA \cite{valentini2022predicting,manocha2021noresqa,hu23d_interspeech,hu24d_interspeech} has gradually gained attention. 
Many preference-score-based SQA methods are derived from MOS-based methods, such as UTP \cite{hu23d_interspeech}, and UTP2 \cite{hu24d_interspeech}. 
However, their focus is often on assisting system-level MOS prediction rather than the accuracy of preference prediction for speech pairs, and there is limited research specifically targeting preference score prediction.

The scarcity of preference data is a major factor hindering the development of preference-score-based SQA methods. 
In recent years, MOS datasets have been constructed from various types of generated speech \cite{cooper2022generalization,wu2019blizzard,maniati22_interspeech}, leading researchers to construct preference data from MOS data to save the cost of obtaining real preference data. 
However, the current methods of constructing preference data also have certain issues. 
Many works \cite{manocha2021noresqa,hu23d_interspeech,hu24d_interspeech}, for the sake of convenience, construct content-mismatched speech pairs from MOS datasets, which fail to cover all possible training-testing scenarios. 
For example, in many applications of preference listening tests, the compared speech samples often have the same content. 
The prediction performance of existing preference-score-based SQA methods in this scenario is unknown, and their universality is difficult to assess.


To overcome the aforementioned issues, this paper proposes a novel universal preference-score-based pairwise speech quality assessment (UPPSQA) model and constructs various types of paired data from the MOS dataset to validate the model's universality across multiple application scenarios. 
Specifically, UPPSQA first predicts the MOS for each speech sample in a pair using a semantic-acoustic-driven MOS prediction (SA-MOS) model, and then calculates the preference score through a preference function. 
We conducted extensive experiments under various training and testing scenarios, including: 1) four types of data combination, where four training-testing datasets are constructed based on whether the paired speech samples have matched content; 2) two types of training label data, i.e., whether MOS labels can be utilized; and 3) two testing methods, i.e., using both in-domain and out-of-domain data for testing. 
Experimental results confirm that the proposed UPPSQA is universal, with its preference prediction accuracy outperforming to the baseline models in most of the above scenarios.

\begin{figure}[t]
    \centering
    \includegraphics[width=0.96\linewidth]{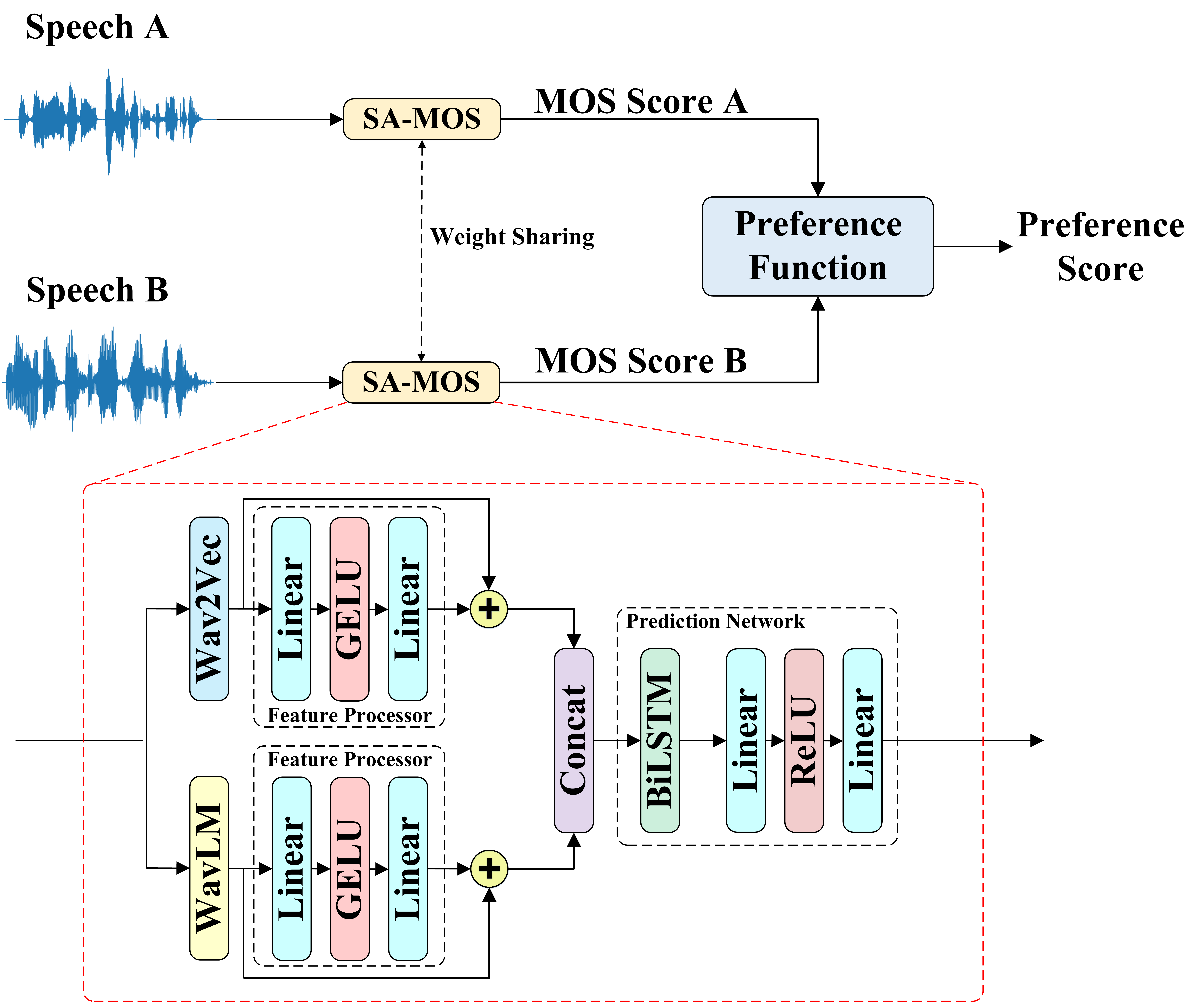}
    \caption{The architecture of the proposed UPPSQA.}
    \label{fig1}
\end{figure}
\vspace{-1.5mm}
\section{Proposed Method} 
\vspace{-1.5mm}
In this section, we will first introduce the architecture of UPPSQA, followed by the training approach and dataset construction method for UPPSQA.

\vspace{-1.5mm}
\subsection{Architecture} 
\vspace{-1.5mm}
Figure \ref{fig1} illustrates the architecture of the proposed UPPSQA. 
We used an identical SA-MOS model to respectively score two speech samples in a pair and then fed the two absolute predicted MOS scores into a preference function to obtain the final predicted preference score. 
\vspace{-1.5mm}
\subsubsection{MOS Prediction for Paired Speech}

As shown in Figure \ref{fig1}, we propose a novel SA-MOS model, which predicts the absolute MOS score for each speech sample in a pair separately. 
SA-MOS separately uses wav2vec2 2.0 \cite{baevski2020wav2vec} and WavLM \cite{chen2022wavlm} to extract semantic and acoustic features from the speech samples to be evaluated, performing a comprehensive assessment that fully leverages the multi-dimensional characteristics of the speech.
Wav2vec 2.0, widely used in various speech tasks such as automatic speech recognition (ASR), is pre-trained on hundreds to thousands of hours of speech data. 
Given its robust semantic representation capabilities, we utilized it as a module for extracting semantic representations. 
WavLM, on the other hand, is pre-trained by mixing signals from multiple speakers but only predicting targets related to the original speaker \cite{chen2022wavlm}. 
This unique pre-training approach enables it to better capture acoustic features. 
We take a weighted sum of the hidden states from all layers of WavLM as the acoustic features, where the weights are learnable. 
Subsequently, the two features are processed deeply by two parallel feature processors, respectively. 
Each feature processor consists of two linear layers and a Gaussian error linear unit (GELU) activation \cite{hendrycks2016gaussian}.
We add the original features and their processed counterparts via residual connections. 
Then we concatenate these two residual connection results along the dimension axis and feed them into a prediction network. 
The prediction network consists of a bidirectional long short-term memory (BiLSTM) layer, followed by a linear layer and a rectified linear unit (ReLU) activation \cite{glorot2011deep}, ultimately yielding the predicted MOS score by a single-node linear layer.

\vspace{-1mm}
\subsubsection{Preference Score Derivation} 
\vspace{-1mm}
 
Suppose $(\bm{x},\bm{y})$ is a speech pair, and the predicted MOS score obtained from SA-MOS for these are $(\hat s_m^{(\bm{x})}, \hat s_m^{(\bm{y})})$. 
Then the UPPSQA use a preference function to derive the preference score $\hat s_p^{(\bm{x},\bm{y})}\in (-1,1)$ of the speech pair, i.e., 
\begin{equation}
    \begin{aligned}
    \hat s_p^{(\bm{x},\bm{y})}=\dfrac{2}{1+e^{-\left(\hat s_m^{(\bm{x})}-\hat s_m^{(\bm{y})}\right)}}-1.
    \end{aligned}
\end{equation}
When $\hat s_p^{(\bm{x},\bm{y})} > 0$, $\bm{x}$ has a higher quality than $\bm{y}$; when $\hat s_p^{(\bm{x},\bm{y})} = 0$, $\bm{x}$ and $\bm{y}$ have equivalent quality; and when $\hat s_p^{(\bm{x},\bm{y})} < 0$, $\bm{x}$ has a lower quality than $\bm{y}$. 

\begin{figure}[t]
    \centering
    \includegraphics[width=1\linewidth]{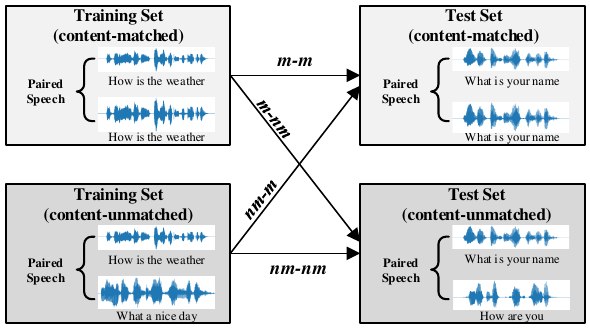}
    \caption{Overview of four training-testing scenarios.}
    \label{fig2}
\end{figure}

\subsection{Training Criteria}

We design two training methods for two different data label scenarios as follows. 
\begin{itemize}
\item {}{\textbf{LA}}: Scenarios where MOS labels are available. 
In this case, we define losses on both MOS and preference scores. 
Assuming the MOS label for $(\bm{x},\bm{y})$ are $(s_m^{(\bm{x})},s_m^{(\bm{y})})$, the MOS loss $\mathcal L_m$ is defined as follows: 
\begin{equation}
    \begin{aligned}
\mathcal L_m=\mathbb E_{(\bm{x},\bm{y})}\left[\left(s_m^{(\bm{x})}-\hat s_m^{(\bm{x})}\right)^2+\left(s_m^{(\bm{y})}-\hat s_m^{(\bm{y})}\right)^2\right].
    \end{aligned}
\end{equation}
The preference score label can be derived from the MOS label, i.e., $s_p^{(\bm{x},\bm{y})}=sgn(s_m^{(\bm{x})}-s_m^{(\bm{y})})$, where $sgn$ is the sign function. 
Then the preference score is defined as follows:
\begin{equation}
    \begin{aligned}
\mathcal L_p=\mathbb E_{(\bm{x},\bm{y})}\left(s_p^{(\bm{x},\bm{y})}-\hat s_p^{(\bm{x},\bm{y})}\right)^2.
    \end{aligned}
\end{equation}
Therefore, the final loss is $\mathcal L=\mathcal L_m+\mathcal L_p$, and it is used to train the SA-MOS model. 

\item {}{\textbf{LM}}: Scenarios where MOS labels are missing. 
This scenario is for preference data that is obtained directly, rather than derived from MOS data. 
In this case, we train the SA-MOS model using only the loss $\mathcal L_p$. 
Since there are no MOS label supervision, SA-MOS effectively becomes an absolute score predictor, rather than a MOS predictor.

\end{itemize}



\subsection{Paired Data Generation}
\label{pair generation} 
As mentioned in Section \ref{Introduction}, the preference score prediction task faces the issue of data scarcity. 
Therefore, we propose a method to construct a dedicated preference score dataset based on the easily accessible MOS dataset. 
Each utterance in the dataset is a five-tuple $(\bm{x},\bm{y},s_m^{(\bm{x})},s_m^{(\bm{y})},s_p^{(\bm{x},\bm{y})})$, including paired speech, paired labeled absolute MOS scores and relative  preference score. 
Based on the practical application scenarios of preference listening tests, we employed two manners for constructing speech pairs. 

\begin{itemize}
\item {}{\textbf{Content-matched approach}}: 
We first perform automatic speech recognition (ASR) on the dataset and then cluster the speech samples based on their textual content. 
Within each cluster, we construct all possible speech pairs. 
In this way, the content of each speech pair is matched.

\item {}{\textbf{Content-unmatched approach}}: 
We first divide the dataset into $K$ groups based on the speech generation system from which the data originates, where $K$ represents the number of systems. 
Then, we iterate through all possible system pairs and randomly select one speech sample from each of the two systems in a system pair to form a speech pair. 
Therefore, a total of $C_K^2$ speech pairs will be formed. 
In this way, the content of most speech pairs is unmatched.

\end{itemize}


As shown in Figure \ref{fig2}, to cover all possible application cases, we constructs four training-testing scenarios to validate the universality of the model, including
\begin{itemize}
\item {}{\textbf{m-m}}: Both training set and test set are content-matched.

\item {}{\textbf{nm-m}}: The training set is content-unmatched while the test set is content-matched. 

\item {}{\textbf{m-nm}}: The training set is content-matched while the test set is content-unmatched. 

\item {}{\textbf{nm-nm}}: Both training set and test set are content-unmatched.

\end{itemize} 


\section{Experimental Setups}
\subsection{Dataset}
\vspace{-1.5mm}
We conducted experiments using two MOS datasets. 
The BVCC dataset \cite{cooper2022generalization} was used for in-domain training and testing, while the BC2019 dataset \cite{wu2019blizzard} was just used for out-of-domain testing. 

The BVCC dataset comprised 7,106 English utterances, officially divided into training, development, and test sets in a ratio of 70$\%$/15$\%$/15$\%$. 
The dataset sources from systems participating in past Blizzard Challenges (BCs), Voice Conversion Challenge systems, and samples generated by ESPNet \cite{watanabe2018espnet}. 
Each utterance in BVCC is rated by 8 listeners, and we utilized the average of these ratings as the ground-truth for absolute MOS scores. 
For the content-matched training and test set constructions, we first used the xlsr-53 ASR model \cite{xu22b_interspeech} for recognition and the DBSCAN clustering algorithm \cite{ester1996density} based on normalized Levenshtein distance for clustering, and then divided the data in the original training set and test set into 317 and 250 same-content classes, respectively. 
Finally, by pairing the samples within each class, we constructed 4,829 speech pairs for the training set and 6,843 speech pairs for the test set, forming the content-matched datasets. 
For the content-unmatched training and test set constructions, we utilized all 175 systems in the original training set and 187 systems in the original test set of BVCC, constructing 15,225 and 17,391 speech pairs, respectively, to form the content-unmatched dataset. 

For BC2019 dataset, we only adopted its test set which comprised 540 Mandarin utterances from the BC 2019 challenge for out-of-domain evaluation. 
Each utterance is scored by 10 to 17 listeners, and we also took the average MOS score as the ground-truth. 
Following the same processing method as the BVCC dataset, we constructed 518 speech pairs for the content-matched test set and 325 speech pairs for the content-unmatched test set.


\subsection{Implementation Details}
\label{Implementation Details}
\vspace{-1.5mm}
In the UPPSQA architecture, we utilized the pretrained Wav2Vec2.0 Base and WavLM Base models available in fairseq to extract semantic and acoustic features for SA-MOS models. 
Both semantic and acoustic features had a dimension of 768. 
The two feature processors in the SA-MOS models consisted of two linear layers with output dimensions of 64 and 768, respectively. 
Additionally, in the prediction network, the BiLSTM layer had 128 nodes and two linear layers had output dimensions of 64 and 1, respectively. 

At the training stage of UPPSQA, we ran the model for 1000 epochs with a batch size of 8, using stochastic gradient descent (SGD) as the optimizer and a learning rate of 0.0001. 
For checkpoint management, we adopted a strategy similar to that used in UTMOS \cite{saeki2022utmos}. 
Specifically, we selected the checkpoint with the highest system-level spearman rank correlation coefficient (SRCC) calculated from the development set according to the absolute score prediction results of the SA-MOS model. 
If the system-level SRCC failed to improve over 15 consecutive epochs, we applied early stopping to prevent overfitting. 
Given that relying on a single checkpoint for metric calculation could introduce variability, we trained the model with five different random seeds, calculated the metric for each resulting checkpoint, and then averaged these metrics to obtain a more robust final result.

\subsection{Baselines and Evaluation Metrics}

Due to the limited open-source work on preference-score-based pairwise SQA, we manually constructed two baselines to compare with our proposed UPPSQA. 
SSL-MOS \cite{cooper2022generalization} is the baseline for the VoiceMOS Challenge 2022. 
It performs MOS prediction by adding a linear layer after Wav2vec2. 
SSL-MOS has been widely adopted in SQA field, so we replaced the SA-MOS in the UPPSQA framework with SSL-MOS as one of the baselines, denoted as SSLSQA.
Additionally, we manually reproduced UTP \cite{hu23d_interspeech}. 
Since we found in experiments that introducing an ID embedding leads to a decrease in model performance, we removed the ID embedding module from UTP. We denote this modified model as UTP$^{\ast}$ and use it as another baseline. 

We calculated the preference score prediction accuracy on the test set for three models, i.e., UPPSQA, SSLSQA, and UTP$^{\ast}$, to compare their performance. 
Assume the test set consists of $N$ speech pairs $(\bm{x}_1,\bm{y}_1),\dots,(\bm{x}_n,\bm{y}_n),\dots,(\bm{x}_N,\bm{y}_N)$. 
If the actual condition of speech pair $(\bm{x}_n,\bm{y}_n)$ is that $\bm{x}_n$ is better than $\bm{y}_n$, and the model also predicts that $\bm{x}_n$ is better than $\bm{y}_n$, the judgment is correct. 
If the model predicts that $\bm{x}_n$ is worse than $\bm{y}_n$ or they are comparable, the judgment is incorrect. 
Suppose that the preference score predicted by the model and the labeled preference score of speech pair $(\bm{x}_n,\bm{y}_n)$ are $\hat s_p^{(\bm{x}_i,\bm{y}_i)}$ and $s_p^{(\bm{x}_i,\bm{y}_i)}$, the preference score prediction accuracy can be calculated as follows:
\begin{equation}
    \begin{aligned}
ACC=1-\dfrac{\sum_{n=1}^N sgn\left(\left|sgn\left(\hat s_p^{(\bm{x}_i,\bm{y}_i)}\right)-s_p^{(\bm{x}_i,\bm{y}_i)} \right| \right)}{N}.
    \end{aligned}
\end{equation}

\section{Experimental Results and Analysis}
\subsection{Comparison with Baselines}

Table \ref{table:in-domain} presents the in-domain experimental results across eight scenarios (i.e., two label conditions $\times$ four training-testing scenarios). 
In most scenarios, our proposed UPPSQA outperformed the two baselines, confirming its universality. 
The UPPSQA outperformed SSLSQA, confirming the effectiveness of combining semantic and acoustic information for scoring. 
The comparison with UTP$^{\ast}$ confirms that our proposed framework is better suited for specialized preference score prediction. 
Next, we compared the performance of UPPSQA under the two label conditions. 
In the same training-testing scenario, the prediction accuracy under the LA condition is significantly higher than that under the LM condition. 
This suggests that introducing the supervision of absolute MOS scores can effectively help the model capture the differences between two paired speech samples, leading to more accurate preference predictions. 
Especially in the m-nm scenario, the accuracy improved by more than 0.1, indicating that MOS supervision effectively compensates for the model's generalization ability on unseen mismatched data. 
Finally, we compared the performance of UPPSQA across different training-testing scenarios. 
Under the LA condition, the models trained with both content-matched and content-unmatched training sets achieved similar accuracy on the same test set. 
Interestingly, under the LM condition, the model trained with the content-unmatched training set achieved noticeably better accuracy. 
This may be attributed to the fact that the content-unmatched training set contained longer speech durations and a greater variety of speech types compared to the content-matched one. 
To some extent, this diversity in the data can also enhance preference score prediction accuracy, especially when MOS labels are unavailable.


\begin{table}[t]
\vspace{-2.0em}
\setlength{\tabcolsep}{3pt}
 \caption{The preference score prediction accuracy of UPPSQA and two baselines under two label conditions and four training-testing scenarios on the in-domain test set.}
 \label{table:in-domain}
 \centering
 \begin{tabular}{c|c|c c c} 
  \hline
  \textbf{Label}&\textbf{Training-Testing}&\multicolumn{3}{c}{\textbf{SQA Methods}} \\
  \cline{3-5}
  \textbf{Condition}&\textbf{Scenario} & \textbf{UPPSQA}& \textbf{SSLSQA} &\textbf{UTP$^{\ast}$}\\
  \hline
  \multirow{4}{*}{LA}&m-m &\textbf{0.790}&0.776&0.771\\  
  &nm-m &\textbf{0.789}&0.784&0.782\\
  &m-nm &\textbf{0.828}&0.817&0.805\\
  &nm-nm &0.823&\textbf{0.825}&0.815\\
  \hline
  \multirow{4}{*}{LM}&m-m &\textbf{0.750}&0.741&0.724\\  
  &nm-m &\textbf{0.780}&0.775&0.760\\
  &m-nm &\textbf{0.720}&0.713&0.710\\
  &nm-nm &\textbf{0.815}&0.810&0.810\\  
  \hline
 \end{tabular}
\end{table}

\begin{table}[t]
\vspace{-0.5em}
\setlength{\tabcolsep}{3pt}
 \caption{The preference score prediction accuracy of UPPSQA and two baselines under two label conditions and four training-testing scenarios on the out-of-domain test set.}
 \label{table:ood}
 \centering
 \begin{tabular}{c|c|c c c} 
  \hline
  \textbf{Label}&\textbf{Training-Testing}&\multicolumn{3}{c}{\textbf{SQA Methods}} \\
  \cline{3-5}
  \textbf{Condition}&\textbf{Scenario} & \textbf{UPPSQA}& \textbf{SSLSQA} &\textbf{UTP$^{\ast}$}\\
  \hline
  \multirow{4}{*}{LA}&m-m &\textbf{0.705}&0.687&0.696\\  
  &nm-m &0.680&\textbf{0.681}&0.670\\
  &m-nm &\textbf{0.748}&0.726&0.745\\
  &nm-nm &\textbf{0.723}&0.714&0.706\\
  \hline
  \multirow{4}{*}{LM}&m-m &\textbf{0.654}&0.564&0.623\\  
  &nm-m &\textbf{0.695}&0.689&\textbf{0.695}\\
  &m-nm &0.594&0.594&\textbf{0.606}\\
  &nm-nm &\textbf{0.714}&0.702&0.712\\  
  \hline
 \end{tabular}
\end{table}

\begin{table}[t]
\vspace{-2.0em}
\setlength{\tabcolsep}{5pt}
 \caption{The preference score prediction accuracy of UPPSQA under two label conditions and solely MOS-based methods at nm-nm training-testing scenario.}
 \label{table:in-domian MOS}
 \centering
 \begin{tabular}{c|c c c} 
  \hline
  &In-Domain&Out-of-Domain \\
  \hline
  UPPSQA (LA) & \textbf{0.823} & \textbf{0.723} \\
  UPPSQA (LM) & 0.815 & 0.714 \\
  Solely MOS-based & 0.819 & 0.695 \\
  \hline
 \end{tabular}
\end{table}

\begin{figure}[t]
\centering
\includegraphics[scale=0.2]{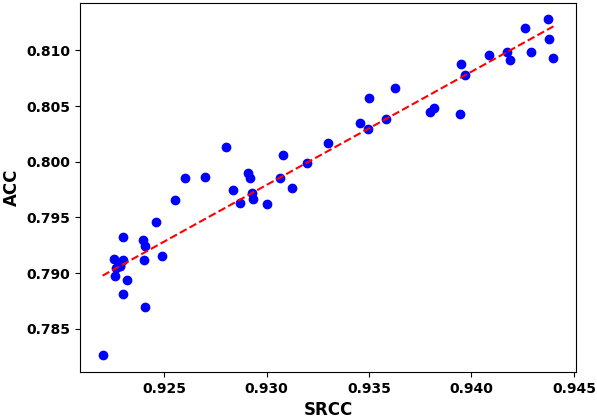}
\caption{The scatter plot showing the correlation between the accuracy (ACC) and the system-level SRCC on the development set for each checkpoint during the training with content-unmatched data under the LA label condition for UPPSQA.}
\label{fig3}
\end{figure}

The out-of-domain experimental results are listed in Table \ref{table:ood}. 
The accuracy on the out-of-domain test set shows a significant drop compared to the in-domain dataset, which is reasonable, as this test set is completely unseen and contains different language content. 
Nevertheless, the proposed UPPSQA still maintained stable performance, with accuracy higher than the baseline in most scenarios. 
The out-of-domain experimental results further confirm that the proposed UPPSQA is universal and can maintain stable performance across various scenarios.

\vspace{-1mm}
\subsection{Analysis and Discussion}
\vspace{-1mm}
\subsubsection{Importance of Pairwise Assessment}

The UPPSQA implements preference score prediction using a pairwise assessment approach. 
However, single assessment approach is more straightforward. 
In this approach, an SA-MOS model is trained using single speech samples, and during testing, the model respectively calculates the MOS scores for the two compared speech samples, and then the preference score is obtained through the preference function. 
Only the MOS loss is defined for training the SA-MOS model. 
We compared UPPSQA with this solely MOS-based approach. 
For simplicity, the experiment is conducted only in the nm-nm scenario. 
When training the SA-MOS model in this approach, the training set with content-unmatched paired samples is transformed into a training set with non-repeating single samples. 
The experimental results are shown in Table \ref{table:in-domian MOS}. 
It is clear that, especially in the out-of-domain test scenario, the solely MOS-based approach exhibited worse prediction accuracy. 
This validates the effectiveness of the pairwise assessment approach proposed in this paper.


\vspace{-1mm}
\subsubsection{Discussion on Model Selection Strategy}
\vspace{-1mm}

As described in Section \ref{Implementation Details}, UPPSQA selects the model based on the system-level SRCC metric on the development set, rather than the preference score accuracy. 
The reason is that calculating accuracy is highly time-consuming, which limits the training efficiency of the model. For example, with content-unmatched data, calculating the SRCC requires running the SA-MOS model 1,066 times, while calculating accuracy requires running it 32,580 times, resulting in a 30-fold reduction in validation speed. 
Figure \ref{fig3} shows the SRCC and accuracy for each checkpoint during training process. 
We can see that they are almost linearly correlated. 
Therefore, using SRCC to select the model on the development set is reasonable and helps improve training efficiency.


\vspace{-2mm}
\section{Conclusion}
\vspace{-1mm}
This paper proposed UPPSQA, a universal preference-score-based pairwise speech quality assessment model. 
UPPSQA uses a semantic-acoustic-driven SA-MOS model to predict the absolute scores of paired speech samples, and then outputs the relative preference scores through a preference function. 
We designed a total of 16 application scenarios based on the MOS datasets to validate the universality of UPPSQA. 
Further improving the prediction accuracy of UPPSQA will be the focus of our future work.

\bibliographystyle{IEEEtran}
\bibliography{mybib}

\begin{thebibliography}{10}
\providecommand{\url}[1]{#1}
\csname url@samestyle\endcsname
\providecommand{\newblock}{\relax}
\providecommand{\bibinfo}[2]{#2}
\providecommand{\BIBentrySTDinterwordspacing}{\spaceskip=0pt\relax}
\providecommand{\BIBentryALTinterwordstretchfactor}{4}
\providecommand{\BIBentryALTinterwordspacing}{\spaceskip=\fontdimen2\font plus
\BIBentryALTinterwordstretchfactor\fontdimen3\font minus \fontdimen4\font\relax}
\providecommand{\BIBforeignlanguage}[2]{{%
\expandafter\ifx\csname l@#1\endcsname\relax
\typeout{** WARNING: IEEEtran.bst: No hyphenation pattern has been}%
\typeout{** loaded for the language `#1'. Using the pattern for}%
\typeout{** the default language instead.}%
\else
\language=\csname l@#1\endcsname
\fi
#2}}
\providecommand{\BIBdecl}{\relax}
\BIBdecl

\bibitem{lo2019mosnet}
C.-C. Lo, S.-W. Fu, W.-C. Huang, X.~Wang, J.~Yamagishi, Y.~Tsao, and H.-M. Wang, ``{MOSNet}: Deep learning-based objective assessment for voice conversion,'' in \emph{Proc. Interspeech}, 2019, pp. 1541--1545.

\bibitem{huang2022ldnet}
W.-C. Huang, E.~Cooper, J.~Yamagishi, and T.~Toda, ``{LDNet}: Unified listener dependent modeling in mos prediction for synthetic speech,'' in \emph{Proc. ICASSP}, 2022, pp. 896--900.

\bibitem{cooper2022generalization}
E.~Cooper, W.-C. Huang, T.~Toda, and J.~Yamagishi, ``Generalization ability of mos prediction networks,'' in \emph{Proc. ICASSP}, 2022, pp. 8442--8446.

\bibitem{saeki2022utmos}
T.~Saeki, D.~Xin, W.~Nakata, T.~Koriyama, S.~Takamichi, and H.~Saruwatari, ``{UTMOS: UTokyo-SaruLab} system for {VoiceMOS Challenge} 2022,'' in \emph{Proc. Interspeech}, 2022, pp. 4521--4525.

\bibitem{shi2024samos}
Y.-F. Shi, Y.~Ai, Y.-X. Lu, H.-P. Du, and Z.-H. Ling, ``{SAMOS: A} neural {MOS} prediction model leveraging semantic representations and acoustic features,'' in \emph{Proc. ISCSLP}, 2024, pp. 199--203.

\bibitem{shah2014better}
N.~B. Shah, S.~Balakrishnan, J.~Bradley, A.~Parekh, K.~Ramchandran, and M.~Wainwright, ``When is it better to compare than to score?'' \emph{arXiv preprint arXiv:1406.6618}, 2014.

\bibitem{vazquez2002reliability}
Y.~Vazquez-Alvarez and M.~Huckvale, ``The reliability of the itu-p. 85 standard for the evaluation of text-to-speech systems,'' in \emph{Proc. ISCSLP}, 2002, pp. 329--332.

\bibitem{kiritchenko2017best}
S.~Kiritchenko and S.~Mohammad, ``Best-worst scaling more reliable than rating scales: {A} case study on sentiment intensity annotation,'' in \emph{Proc. ACL}, 2017, pp. 465--470.

\bibitem{valentini2022predicting}
C.~Valentini-Botinhao, M.~S. Ribeiro, O.~Watts, K.~Richmond, and G.~E. Henter, ``Predicting pairwise preferences between {TTS} audio stimuli using parallel ratings data and anti-symmetric twin neural networks,'' in \emph{Proc. Interspeech}, 2022, pp. 471--475.

\bibitem{manocha2021noresqa}
P.~Manocha, B.~Xu, and A.~Kumar, ``{NORESQA: A} framework for speech quality assessment using non-matching references,'' \emph{Advances in neural information processing systems}, vol.~34, pp. 22\,363--22\,378, 2021.

\bibitem{hu23d_interspeech}
C.-H. Hu, Y.~Yasuda, and T.~Toda, ``Preference-based training framework for automatic speech quality assessment using deep neural network,'' in \emph{Proc. Interspeech}, 2023, pp. 546--550.

\bibitem{hu24d_interspeech}
C.~Hu, Y.~Yasuda, and T.~Toda, ``Embedding learning for preference-based speech quality assessment,'' in \emph{Proc. Interspeech}, 2024, pp. 2685--2689.

\bibitem{wu2019blizzard}
Z.~Wu, Z.~Xie, and S.~King, ``The {Blizzard Challenge} 2019,'' in \emph{Proc. Blizzard Challenge Workshop}, 2019.

\bibitem{maniati22_interspeech}
G.~Maniati, A.~Vioni, N.~Ellinas, K.~Nikitaras, K.~Klapsas, J.~S. Sung, G.~Jho, A.~Chalamandaris, and P.~Tsiakoulis, ``{SOMOS: The Samsung Open MOS Dataset for the Evaluation of Neural Text-to-Speech Synthesis},'' in \emph{Proc. Interspeech}, 2022, pp. 2388--2392.

\bibitem{baevski2020wav2vec}
A.~Baevski, Y.~Zhou, A.~Mohamed, and M.~Auli, ``wav2vec 2.0: {A} framework for self-supervised learning of speech representations,'' \emph{Advances in neural information processing systems}, vol.~33, pp. 12\,449--12\,460, 2020.

\bibitem{chen2022wavlm}
S.~Chen, C.~Wang, Z.~Chen, Y.~Wu, S.~Liu, Z.~Chen, J.~Li, N.~Kanda, T.~Yoshioka, X.~Xiao \emph{et~al.}, ``{Wavlm: L}arge-scale self-supervised pre-training for full stack speech processing,'' \emph{IEEE Journal of Selected Topics in Signal Processing}, vol.~16, no.~6, pp. 1505--1518, 2022.

\bibitem{hendrycks2016gaussian}
D.~Hendrycks and K.~Gimpel, ``Gaussian error linear units (gelus),'' \emph{arXiv preprint arXiv:1606.08415}, 2016.

\bibitem{glorot2011deep}
X.~Glorot, A.~Bordes, and Y.~Bengio, ``Deep sparse rectifier neural networks,'' in \emph{Proceedings of the fourteenth international conference on artificial intelligence and statistics}.\hskip 1em plus 0.5em minus 0.4em\relax JMLR Workshop and Conference Proceedings, 2011, pp. 315--323.

\bibitem{watanabe2018espnet}
S.~Watanabe, T.~Hori, S.~Karita, T.~Hayashi, J.~Nishitoba, Y.~Unno, N.~E.~Y. Soplin, J.~Heymann, M.~Wiesner, N.~Chen \emph{et~al.}, ``{ESPNet}: End-to-end speech processing toolkit,'' in \emph{Proc. Interspeech}, 2018, pp. 2207--2211.

\bibitem{xu22b_interspeech}
Q.~Xu, A.~Baevski, and M.~Auli, ``{Simple and Effective Zero-shot Cross-lingual Phoneme Recognition},'' in \emph{Proc. Interspeech}, 2022, pp. 2113--2117.

\bibitem{ester1996density}
M.~Ester, H.-P. Kriegel, J.~Sander, X.~Xu \emph{et~al.}, ``A density-based algorithm for discovering clusters in large spatial databases with noise,'' in \emph{kdd}, vol.~96, no.~34, 1996, pp. 226--231.

\end{thebibliography}

\end{document}